\documentclass{aa}

\usepackage{graphicx}
\usepackage{txfonts}
\usepackage{natbib}
\bibpunct{(}{)}{;}{a}{}{,}

\begin{document}

   \title{Cold fronts and multi-temperature structures in the core of\\ Abell 2052}

   \author{J. de Plaa\inst{1} \and
          N. Werner\inst{2} \and
	  A. Simionescu\inst{2} \and
	  J. S. Kaastra\inst{1,3} \and
	  Y. G. Grange\inst{1} \and
	  J. Vink\inst{3} 
	  }

   \institute{SRON Netherlands Institute for Space Research,
              Sorbonnelaan 2, 3584 CA Utrecht, The Netherlands\\
              \email{j.de.plaa@sron.nl}
         \and
	      Kavli Institute for Particle Astrophysics and Cosmology,
	      Stanford University, 382 via Pueblo Mall, Stanford, CA 94305-4060, USA
	 \and 
              Astronomical Institute, Utrecht University,
	      PO Box 80000, 3508 TA Utrecht, The Netherlands
             }

   \date{Received 11 June 2010; accepted 13 August 2010}

   \abstract
   {The physics of the coolest phases in the hot Intra-Cluster Medium (ICM) of clusters of galaxies is 
   yet to be fully unveiled. X-ray cavities blown by the central Active Galactic Nucleus (AGN) contain enough energy to
   heat the surrounding gas and stop cooling, but locally blobs or filaments of gas appear to be able to 
   cool to low temperatures of 10$^4$ K. In X-rays, however, gas with temperatures lower than 0.5 keV is not observed.}
   {We aim to find spatial and multi-temperature structures in the hot gas of the cooling-core cluster Abell 2052 that 
   contain clues on the physics involved in the heating and cooling of the plasma. }
   {2D maps of the temperature, entropy, and iron abundance are derived from XMM-Newton
   data of Abell 2052. For the spectral fitting, we use Differential
   Emission Measure (DEM) models to account for the multi-temperature structure.}
   {About 130 kpc South-West of the central galaxy, we discover a discontinuity in the surface brightness of
   the hot gas which is consistent with a cold front. Interestingly, the iron abundance jumps from $\sim$0.75 to $\sim$0.5 
   across the front. 
   In a smaller region to the North-West of the central galaxy we find a relatively high contribution of cool 
   0.5 keV gas, but no X-ray emitting gas is detected below that temperature. However, the region appears to be associated 
   with much cooler H$\alpha$ filaments in the optical waveband.}
   {The elliptical shape of the cold front in the SW of the cluster suggests that the front is
   caused by sloshing of the hot gas in the clusters gravitational potential. This effect
   is probably an important mechanism to transport metals from the core region to the outer parts 
   of the cluster. The smooth temperature profile across the sharp jump in the metalicity indicates the presence 
   of heat conduction and the lack of mixing across the discontinuity.  
   The cool blob of gas NW of the central galaxy was probably pushed away from the core and 
   squeezed by the adjacent bubble, where it can cool efficiently and relatively undisturbed by the AGN.
   Shock induced mixing between the two phases may cause the 0.5 keV gas to cool non-radiatively and explain
   our non-detection of gas below 0.5 keV.}

   
   \maketitle

\section{Introduction}

Our understanding of heating and cooling mechanisms operating in the hot Intra-Cluster Medium (ICM) in the 
cores of clusters of galaxies is not yet complete. The cooling times of the relatively 
dense X-ray emitting plasma in the central region are short compared to the 
Hubble time, which lead to the theory of cooling-flows \citep[e.g.][]{fabian1994}. 
The first cluster observations with the high-resolution Reflection Grating Spectrometer \citep[RGS,][]{herder2001} 
aboard XMM-Newton show, however, that the amount of cool gas in the centre of 
clusters is much smaller than expected \citep{peterson2001,tamura2001a,kaastra2001,peterson2003}. 
In recent years, it has been suggested that this lack of cool gas can be explained by feedback
from the central Active Galactic Nucleus (AGN). The relativistic plasma in the jets originating 
from this accreting central super-massive black hole creates cavities in the hot X-ray emitting gas, 
which appear as dark regions in X-ray images of clusters. The energy enclosed in these 
plasma bubbles appears to be enough to balance the cooling flow \citep[e.g.][]{churazov2002,bruggen2002,birzan2004}. 
The mechanism responsible for gently transferring the energy from the bubbles to the X-ray gas is, 
however, still unclear.

The lowest temperatures detected in clusters through X-rays are about 0.5 keV, which 
appears to be a universal low-temperature floor for clusters and groups. However, these 
low-temperature regions are usually found in clusters which show AGN activity, like, 
for example, \object{Hydra A} \citep{mcnamara2000}, \object{Perseus} \citep{sanders2007}, and \object{M87} 
\citep[e.g.][]{werner2010}. This appears contradictory, 
because AGN are supposed to heat the gas. Studies of the volume filling fraction 
of this 0.5 keV component \citep[e.g.][]{sanders2002,sanders2004} show that the 
cool gas is actually distributed in clumps or filaments. This suggests that gas 
can cool locally to low temperatures despite the fact
that it is embedded in hotter gas. In many clusters these cool regions
are seen at the same position as bright H$\alpha$ regions and filaments detected in optical 
images of these clusters. The connection between the 0.5 keV gas and H$\alpha$ emission
is an important piece of the puzzle of understanding heating and cooling in cluster cores.      

On slightly larger scales, disturbances due to merger events cause the dark matter and 
subsequently the hot X-ray gas to oscillate in the deep gravitational potential well of a cluster.
This sloshing of gas is usually recognised in X-ray images through asymmetries and jumps in the
surface brightness of the hot gas. The underlying density discontinuity, also called a cold front, 
marks the boundary between relatively cool moving gas from the central part of the cluster 
and the hot gas in the outer regions \citep[see][for a review]{markevitch2007}. The slow 
release of mechanical energy from the sloshing movements and the mixing of hot gas from 
the outer parts into the cooling core may both contribute to heat the inner parts of the 
cluster in concurrence with the AGN. Sloshing is probably also an important mechanism for 
transporting metals from the metal-rich core to the metal-poor outer parts \citep{simionescu2010}. 

The cluster of galaxies \object{Abell 2052} is also a bright cool-core cluster in X-rays showing 
AGN activity and H$\alpha$ regions \citep{blanton2001}. The cluster was detected and studied 
in the 1970's \citep{giacconi1972,heinz1974}.
\object{Abell 2052} is also extensively studied with the current generation of X-ray observatories.
It was member of several samples of clusters observed with ASCA \citep{finoguenov2001} 
and XMM-Newton \citep[e.g.][]{kaastra2004,tamura2004,deplaa2007}. 
Chandra images \citep{blanton2001,blanton2003,blanton2009} show evidence of AGN feedback 
by the central radio source \object{3C 317} \citep{zhao1993}. In the inner core ($<$ 1.0$^{\prime}$) 
of the cluster, the X-ray image shows bubbles which are associated with 
the radio lobes of \object{3C 317}. The energy contained in the cavities is thought to effectively heat 
the Intra-Cluster Medium (ICM). In addition, density discontinuities, probably shocks
have been detected by Chandra \citep{blanton2009} just outside the region with the bubbles.

We have obtained a long observation of Abell 2052 with XMM-Newton, which was 
performed in 2007. In this paper, we combine this new observation with an older AO1 
observation of $\sim$ 40 ks and study the thermodynamics in and around the core. 
We use the spatially resolved spectra from the European Photon Imaging Camera (EPIC) 
to make two-dimensional maps.

In our analysis, we use H$_{0}$ = 70 km s$^{-1}$ Mpc$^{-1}$, $\Omega_{\mathrm{m}}$ = 0.3, 
and $\Omega_{\Lambda}$ = 0.7. At the redshift of Abell 2052 ($z$=0.0348), an angular 
distance of 1$^{\prime}$ corresponds to 42 kpc. The elemental abundances presented 
in this paper are given relative to the proto-solar abundances from \citet{lodders2003}. 
Measurement errors are given at 68\% confidence level.

\section{Data analysis}

Abell 2052 has been observed with XMM-Newton in two observation campaigns. 
Two observations have been performed in 2000 with a total exposure time
of 37 ks and another ten exposures were obtained in 2007 with a total 
observing time of 221 ks. 

The event files for the MOS and pn detectors are re-produced using the standard
SAS 9.0.0 pipeline tools. In order to reduce the soft-proton background, we filter 
the data with a 10--12 keV count-rate threshold. We determine
the threshold for the 2000 (AO1) and 2007 (AO4) data separately. For the AO4 data
we combined all the separate observations to obtain one 10-12 keV light curve per 
instrument. A count-rate histogram created from this light curve with 100 s wide bins 
is fitted with a Gaussian to determine the count rate of the quiescent emission ($N$). We then define
the minimum and maximum count rate ($T$) to be $T = N~\pm~3\sqrt{N}$, which is the 3$\sigma$
deviation from the Poissonian mean. We apply this to both the AO1 and AO4 data sets. 
For the AO1 data, we obtain allowed count rates 
in the range 0.01--0.18 (MOS1), 0.01--0.18 (MOS2), and 0.11--0.43 (pn) in units counts s$^{-1}$.
The average rate in counts s$^{-1}$ in the AO4 data was somewhat higher: 0.03--0.27 (MOS1), 
0.04--0.28 (MOS2), and 0.28--0.70 (pn). This is due to the secular evolution of the XMM-Newton 
instrumental background.   
 
In Table~\ref{tab:exposure}, we list all used observations of \object{Abell 2052} with their 
respective exposure times after filtering for flares. Two other observations are discarded, 
because the count rate was above the threshold during the entire exposure. 
Only eight of the brightest point sources in the field were selected 
by eye using a combined image of MOS and pn. Regions with a radius of 15$^{\prime\prime}$ 
around these sources were subsequently excluded from the data set.

\begin{table}[t]
\caption{List of observation IDs used for the analysis with their useful 
exposure time in ks after flare filtering.}
\label{tab:exposure}
\begin{center}
\begin{tabular}{lrrr}
\hline\hline
ObsID		&MOS1 	&MOS2	&pn   \\
\hline
0109920101 	& 29.8& 29.9 & 22.8   \\
0109920301	&  2.4&  2.4 & -      \\
0401520501 	&  3.9&  4.4 &  4.5   \\
0401520601 	&  5.1&  5.3 &  2.5   \\
0401520801 	&  9.8& 10.3 &  2.3   \\
0401520901 	&  8.2&  8.4 &  2.2   \\
0401521101 	&  8.0&  7.9 &  3.6   \\
0401521201 	& 18.9& 18.0 & 12.1   \\
0401521601 	&  8.2&  8.8 &  2.7   \\
\hline
Total		& 94.3& 95.4 & 52.7   \\
\hline
\end{tabular}
\end{center}
\end{table}

\subsection{Creating maps}

In order to resolve structures in temperature and iron abundance,
we divide the data in small regions from which spectra can be extracted.  
To obtain sufficient signal-to-noise per spatial bin, we use the Weighted Voronoi Tessellation (WVT) binning 
algorithm by \citet{diehl2006}, which is a generalization of the \citet{cappellari2003} 
Voronoi binning algorithm. We apply the binning to the total background 
subtracted image and create maps with a signal-to-noise of 150$\sigma$ per bin. 
Because of the relatively low surface brightness of the source 
compared to the X-ray background in the outer parts of the cluster, we only select 
bins within a radius of 4$^{\prime}$ ($\sim$ 168 kpc) around the centre of the central cD galaxy. 

For every bin, we extract the event files and spectra. Then, we calculate the
response matrices and effective area files for all spatial bins using one AO1 pointing and one 
AO4 pointing. These response and effective area files are also used for the other pointings 
of the same AO in order to reduce computing time. We checked that the response is 
stable enough within one AO. Also, the pointings within one AO were very similar, causing the
extraction regions defined in WCS coordinates to cover the same area in detector coordinates 
for every pointing. Therefore, using one response matrix per spatial bin per AO is justified.
As we are mainly interested in relative differences between bins, 
we adopt a relatively simple background treatment for the maps. We subtract a scaled 
filter wheel closed spectrum from the source spectrum. The scaling factor is based 
on the out-of-field-of-view events in MOS. The spectra for each bin are fitted 
simultaneously using the SPEX spectral fitting package. The Cosmic X-ray Background 
(CXB) is included as a set of model components in the fit. For the $N_{\mathrm{H}}$ 
we use the Galactic value of 2.71$\times$10$^{20}$ cm$^{-2}$ \citep{kalberla2005}.

\subsection{Fitting interesting regions}

Apart from spectra extracted from binned maps, we also carefully 
analyse spectra extracted from certain interesting regions that we identify in the 
maps. In addition, the parameters of the CXB components need to be estimated, 
because they are needed for fitting the maps. 
Therefore, we use a separate procedure for these high signal-to-noise spectra. 
The background treatment is very important when fitting spectra extracted from
the outer regions of a cluster, where the flux of the cluster emission is
comparable to the background. 

Correcting for all the various background components in XMM-Newton data is 
very challenging, because many components either depend on the position on the
sky or depend on the epoch of observation. The fact that we have obtained a lot of 
relatively short exposures with different background conditions, demands a 
careful treatment of all the components. Although there are very well constructed
methods like, for example, \citet{snowden2008}, there are a number of drawbacks
when they are applied to this data set. The success of these methods strongly depends on
the events registered outside the XMM-Newton field of view. Because the pn instrument 
has very small out-of-field-of-view regions, their method is practically unusable for 
pn data. Secondly, the amount of counts in the out-of-field of view regions is rather 
small for short exposure times, which puts a relatively large uncertainty on the 
derived scaling factor for the instrumental background. This drawback can be avoided by
stacking event files, but only if the gain and background were similar during the 
observations. Here, we present a method 
in which we model all the background components during spectral fitting without 
subtracting any background spectrum beforehand. This allows us to use both EPIC MOS and pn data. 

\begin{table}[t]
\caption{Modeled instrumental lines for EPIC MOS and pn.}
\begin{center}
\begin{tabular}{lrrr}
\hline\hline
Element		& Energy	& MOS	& pn \\
		& (keV)		&	&    \\
\hline
Al K$\alpha$	& 1.486 	& +	& + \\
Al K$\beta$	& 1.557 	& +	& + \\
Si K$\alpha$	& 1.740 	& +	& - \\
Si K$\beta$	& 1.835 	& +	& - \\
Ti K$\alpha$	& 4.51		& -	& + \\
Cr K$\alpha$	& 5.41 		& +	& + \\
Mn K$\alpha$	& 5.89 		& +	& - \\
Fe K$\alpha$	& 6.40 		& +	& - \\
Ni K$\alpha$	& 7.47  	& -	& + \\
Cu K$\alpha$ 	& 8.04  	& - 	& + \\
Zn K$\alpha$	& 8.63 		& +	& + \\
Cu K$\beta$	& 8.90  	& -	& + \\
Zn K$\beta$	& 9.57  	& -	& + \\
Au K$\alpha$	& 9.72 		& +	& + \\
\hline
\end{tabular}
\end{center}
\label{tab:fluorescence}
\end{table}

The first set of background components that we model is the instrumental background due
to hard particles. These are mostly very energetic particles that enter the detector from all
directions. The total spectrum of these components can be observed when the
`closed' filter has been chosen. The typical spectrum is composed of a broken power law and 
several instrumental fluorescence lines. Since these background events are not caused by 
reflected X-rays from the mirror, the models should not be folded with the effective area (ARF). 
The normalisation of the broken power law can vary in time independently of the lines. 
During the mission, the normalisation 
of the power-law component has slowly increased. We assume that the power-law index does not 
change significantly with time. To find the average parameters of the power law, we fit 
the spectra derived from the merged closed filter observations provided by the XMM-Newton 
Science Operations Centre (SOC).
For every source region that we analyse, we extract a spectrum from a closed-filter region 
with the same detector coordinates. Then, we fit a broken power-law model to the data
including delta lines to model the fluorescence lines (see Table~\ref{tab:fluorescence} for a list of 
lines for MOS and pn). The parameters for the power-law index, the change in index, and
the break energy are used when fitting the source spectrum. In the fit, the normalisations 
of the broken power law and the delta lines are allowed to vary within a bandwidth of a 
factor of two with respect to the best fit closed-filter results. This is to limit the 
possible effect that genuine spectral features are compensated in the fit by increasing 
the normalisation of background components.  
  
The second set of background components is the Cosmic X-ray Background (CXB). The high-energy part 
of this background spectrum is dominated by unresolved point sources, which is described well
by a power-law component with a slope of $\Gamma$ = 1.41 \citep{deluca2004}. We fix the 
normalisation of the power law to their value which corresponds to a 2--10 keV flux of 
2.24 $\times$ 10$^{-14}$ W m$^{-2}$ deg$^{-2}$. In this paper, we do not correct for the 
exclusion of point sources which affects the flux of this component, because we mainly focus 
on the inner 5$^{\prime}$ of the cluster where the influence of small variations in 
this component are negligible. Below 
2 keV, the spectrum consists of contributions from the local-hot bubble and Galactic thermal 
emissions \citep[e.g.][]{kuntz2000,kuntz2008b}. We model these components with two 
single-temperature plasma models assuming Collisional Ionisation Equilibrium (CIE). The coolest
component is associated with the Local Hot Bubble and is not affected by absorption. The temperatures
of the two components can vary slightly over the sky. Therefore, we fit the temperature and 
normalisations of these components in a 9.0$^{\prime}$--12$^{\prime}$ annulus around the cluster
centre in the 0.4--12 keV band. In this annulus, the normalisations and temperatures are well 
constrained. We set the temperatures for the two components to their (rounded) best fit values 
of 0.120 ($\pm$0.004) keV and 0.290 ($\pm$0.006) keV, respectively. The 0.1--2.5 keV flux of 
the 0.12 keV component is (5.03 $\pm$ 0.15) $\times$ 10$^{-14}$ W m$^{-2}$ deg$^{-2}$ and for 
the 0.29 keV component it is (1.87 $\pm$ 0.04) $\times$ 10$^{-14}$ W m$^{-2}$ deg$^{-2}$. We 
assume solar abundances for the Local Hot Bubble and use 0.7 times solar for the more distant 
0.29 keV component, which was a typical value determined from the fits. With this set of models
we have a good parametrisation of the CXB which can be applied to spectra extracted from other 
areas within the field of view.  

There are two other potentially important background components: charge exchange emission from
interactions between the solar wind and solar-system material, and quiescent soft-protons. The 
charge exchange emission consists of time variable line emission below 1 keV. We checked the 
low-energy light curves for variability, but no significant flares were found. Soft-proton flares 
were already removed from the data by rather strict filtering. A low quiescent level of soft-proton 
contamination is probably still present. To get an estimate of the soft-proton flux, we compared
the out-of-field-of-view areas with the count rate in the field of view in the 10--12 keV band.
We did not find a significant difference between the two count rates, which implies that this
component is indeed very weak. Another check would be whether the derived spectral models are
stable with respect to the observation epoch and soft-proton rate. We have fitted the AO1 and AO4 spectra 
extracted from a region in the centre of the cluster separately and find that the derived models parameters, 
like temperature and iron abundance, are consistent within the error bars and within 5\%. 
Therefore, we ignore the soft-proton contribution for now
and let it be partly compensated by the normalisation of the power law of the instrumental background.
An independent measurement of the soft-proton flux would be necessary to include this component in 
the model.

\subsection{Surface brightness profiles}

For diagnostics of density discontinuities it is useful to create profiles of, for example, 
surface brightness. We derive these profiles using images extracted from 
the event files, background images, and exposure maps. Regions are selected using a mask image.
The images are extracted with the same WCS reference coordinates, which means that the pixels 
in every image represent the same area on the sky. We add the source counts, background counts, 
and exposure values for every pixel in all pointings and then we calculate the background 
subtracted count rate. We bin the profile in 0.1$^{\prime}$ bins and calculate the angular distance 
scale with respect to the central cD galaxy. Throughout this paper, we use the coordinates 
$\alpha$ 15$^\mathrm{h}$16$^\mathrm{m}$44.47$^\mathrm{s}$ $\delta$ 7$^{\circ}$01$^{\prime}$18.47$^{\prime\prime}$ for the
centre of the cluster. 

\section{Spectral models}
\label{sec:models}

For spectral fitting we use the publicly available SPEX spectral fitting package
\citep{kaastra1996}. The code contains several methods
to fit multi-temperature Differential Emission Measure (DEM) models to X-ray spectra.
For cluster emission, the Gaussian ({\it gdem}) and power-law ({\it wdem}) parametrisation
have been successfully fitted before. However, from the fits it is usually not possible
to distinguish between these different DEM shapes \citep[e.g.][]{deplaa2006}. 
  
\subsection{GDEM}

The first DEM model that we use is a Gaussian Differential Emission Measure (GDEM) distribution. 
The Gaussian is either defined on a logarithmic temperature grid ($\mathrm{log}~T$) or a linear grid:
\begin{equation}
Y(x) = \frac{Y_0}{\sigma_{\mathrm{T}} \sqrt{2\pi }} \mathrm{e}^{-(x-x_0)^2 / 2\sigma^2_{\mathrm{T}}}.
\end{equation} 
For the logarithmic grid, $x=\mathrm{log}~T$ and $x_0=\mathrm{log}~T_0$ where $T_0$ is the central temperature 
of the distribution. In the linear case, $x=T$ and $x_0=T_0$. 
The width of the Gaussian is $\sigma_{\mathrm{T}}$. 

\subsection{WDEM}

The second empirical model that we use in this paper is known as {\it wdem} 
\citep{kaastra2004,deplaa2005}, which proved to be successful in fitting 
cluster cores \citep[e.g.][]{kaastra2004,werner2006}. This model is 
a DEM model where the differential emission 
measure is distributed as a power law ($\mathrm{d}Y/\mathrm{d}T \propto T^{1/\alpha}$) 
with a high ($T_{\mathrm{max}}$) and low-temperature cut-off ($\beta T_{\mathrm{max}}$)
with $0.1 < \beta < 1.0 $ \citep{deplaa2006}. The statistics do not always allow to fit 
$\alpha$ and $\beta$ simultaneously, because the two parameters are degenerate for spectra with
low statistics. Where the statistics do allow it, we leave $\beta$ free in fits to determine
the low-temperature floor. When the statistics are too low, we fix $\beta$ to 0.1. Since we 
expect temperatures in the range from about 0.5 keV up to 4.0 keV, a value of 0.1 for $\beta$ 
is reasonable lower limit. If this $\beta$ is too low for a certain spectrum, then the fit can  
many times compensate for this by decreasing the value for $\alpha$ without increasing the best 
C-statistics value significantly.   

\subsection{Multi-temperature models}

The last multi-temperature model we use, is not a continuous DEM model, but just a
small grid of four single-temperature (CIE) models \citep{sanders2004}. 
We use two varieties. In the first model, we set the temperatures of the individual components 
to fixed temperatures of 0.5 keV, 1.0 keV, 2.0 keV, and 4.0 keV. For the second model, the temperature
of the highest component is left free and the other three temperature components are coupled with factors
of 0.5, 0.25, and 0.125 respectively.    
In both models, the abundances of individual elements are coupled to the abundance values in the first 
component, and the normalisations of the temperature components are left free.

\section{Results}

\begin{figure}[t]
\includegraphics[width=\columnwidth]{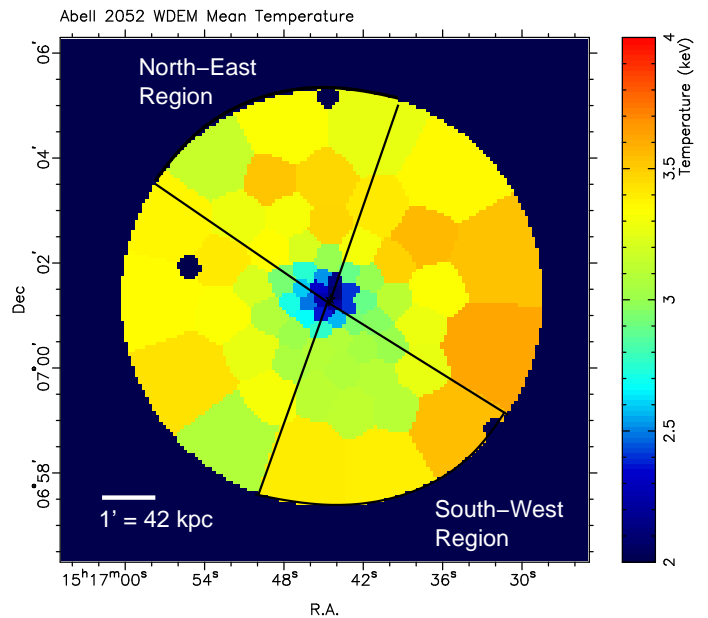}
\caption{Mean temperature map derived from {\it wdem} fits to all spectra extracted from the bins
in the picture above. The temperatures vary from 2 keV in the core up to 3.6 keV in the outer parts
of the cluster. The error on the temperature in the majority of the bins is less than 10\%. 
In the core, a star symbol marks the centre of the main cD galaxy. The areas indicated by the
thick lines are the regions used to compare the radial profiles in the Northern-Eastern and 
South-Western parts of the cluster.}
\label{fig:wdem-tempmap}
\end{figure}

\begin{figure}[t]
\includegraphics[width=\columnwidth]{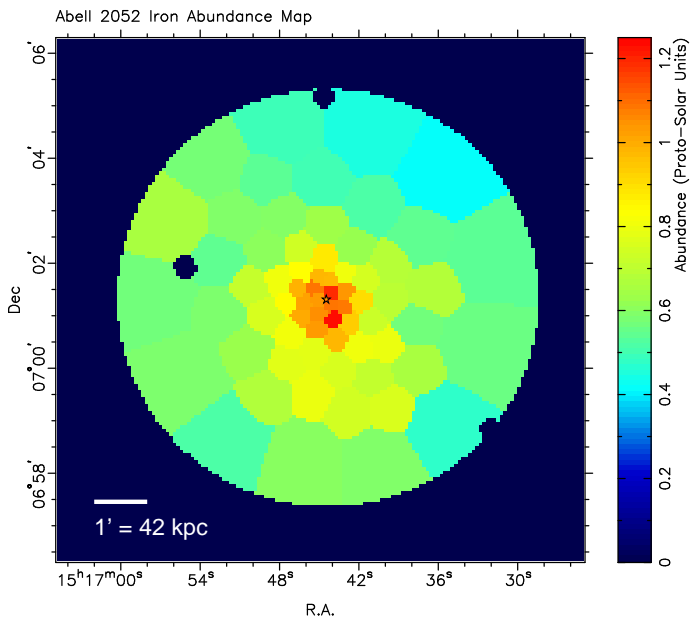}
\caption{Iron abundance map with respect to the proto-solar abundances by \citet{lodders2003}.
Uncertainties on the abundance values range from 5\% in the centre up to 10\% in the outer parts. 
In the core, a star symbol marks the centre of the main cD galaxy.}
\label{fig:wdem-femap}
\end{figure}

\begin{figure}[t]
\includegraphics[width=\columnwidth]{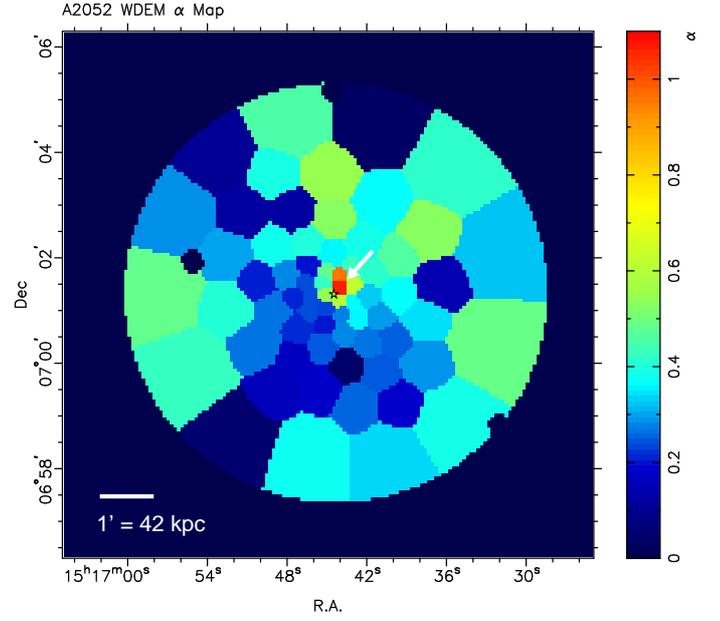}
\caption{Map of the value of the $\alpha$ parameter in the {\it wdem} model with $\beta$ fixed to 0.1. 
The higher the $\alpha$ parameter, the higher the relative contribution of cooler components in the spectrum. 
The error on the $\alpha$ parameter is less than 30\% across the image, except for the bins where the value 
of $\alpha$ is nearly zero. These bins are within 2$\sigma$ significance consistent with their neighbours 
which have higher $\alpha$ values. In the central region, the typical uncertainty is less than 10\%.
In the core, a star symbol marks the centre of the main cD galaxy and the white arrow points out the 
bins with the highest $\alpha$ value.}
\label{fig:wdem-alphamap}
\end{figure}

In Figs.~\ref{fig:wdem-tempmap}, \ref{fig:wdem-femap}, \ref{fig:wdem-entropymap}, and \ref{fig:wdem-alphamap}, 
we show maps based on a {\it wdem} fit to the data with the low-temperature cut-off ($\beta$)
fixed to 0.1. A multi-temperature model is needed, 
because a single temperature model does not provide an acceptable fit to the spectra 
extracted near the centre of the cluster. Assuming single-temperature gas, we find typical 
reduced C-statistic values in the central region that are of the order of 3 times the d.o.f. 
Maps using single-temperature models have been created, but they show unphysical structure 
in the pressure map. Therefore, we only show the maps extracted from multi-temperature fits.

The weighted mean temperature of the best-fit {\it wdem} distribution, is shown in 
Fig.~\ref{fig:wdem-tempmap}. The temperature ranges from about 2 keV in the centre of the
cluster up to 3.6 keV at a radius of $\sim$3.5$^{\prime}$. This trend in temperature 
is typical for a cooling-core cluster. Interestingly, the map shows the presence of
an asymmetry in temperature. A large area toward the South-West of the cluster has a 
significantly lower temperature with respect to the gas in the other directions. 
This area is approximately 120 $\times$ 120 kpc in size and the map suggests it extends 
to a distance of about 2.5$^{\prime}$ from the cluster centre.

Also the iron abundance map in Fig.~\ref{fig:wdem-femap} shows a similar asymmetry. Like 
other cooling-core clusters, the iron abundance is peaked in the centre. In the case of \object{Abell 2052}, 
the iron abundance reaches a peak value of about 1.2 times solar in the centre and drops off to about 
0.5 solar at a distance of 4$^{\prime}$ from the core. In the same South-Western region where the map
in Fig.~\ref{fig:wdem-tempmap} shows lower temperatures, the iron abundance is relatively high. In this 
region, typical iron abundances are around 0.75 times solar, while on the opposite side of the cluster they 
are around 0.55 times solar.

\begin{figure}[t]
\includegraphics[width=\columnwidth]{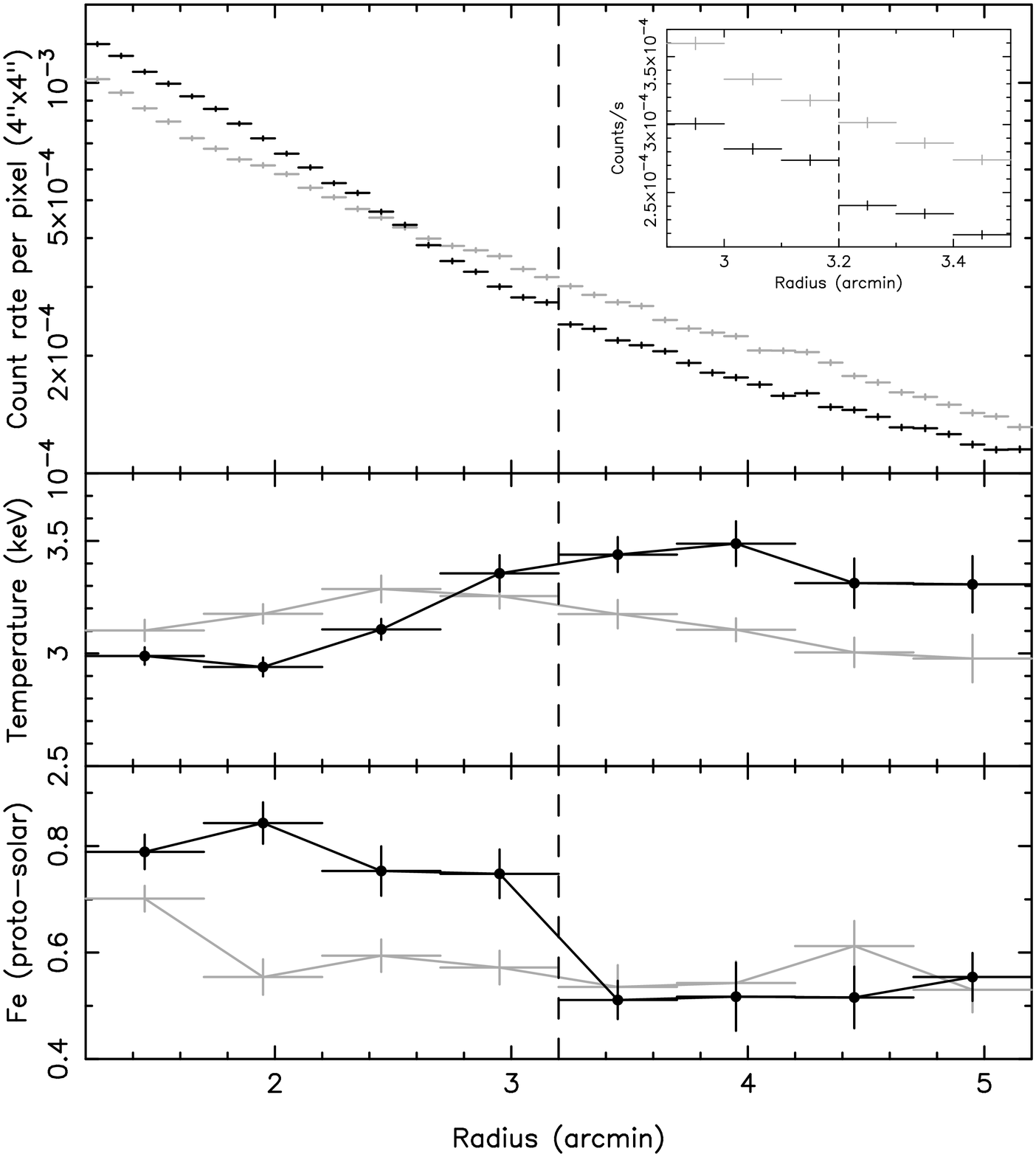}
\caption{Mean count rate per pixel, temperature, and iron abundance profiles extracted from the cool 
South-Western region ({\it black}) and the opposite North-Eastern side ({\it gray}) of Abell 2052. 
The upper panel shows the count rate as a function of the
distance toward the central cD galaxy in the 0.5--10.0 keV band. The inset is a blow-up of the same data
around the surface-brightness discontinuity. The second and third panel show the
profiles of the temperature and iron abundance respectively. The dashed vertical line indicates the position
of the surface brightness discontinuity.}
\label{fig:profiles}
\end{figure}

In order to see whether this area showing lower temperatures and high iron abundance 
is bound by a discontinuity, we extract a profile of the count rate in the direction 
of the asymmetry. The region from which we extract the count rates is the 
South-Western pie shaped region in 
Fig.~\ref{fig:wdem-tempmap}. We bin the count-rates obtained from each image pixel into 
0.1$^{\prime}$ bins and the distance is chosen to be the radial distance toward the central 
cD galaxy. The count-rate profile is shown in the upper panel of Fig.~\ref{fig:profiles}. 
Indeed the profile is showing a jump around 3.2$^{\prime}$. The inset in the upper panel 
of Fig.~\ref{fig:profiles} shows a blow-up of this jump, which is not seen in the count 
rates determined from the opposite side of the cluster, which are shown in gray.

More detailed properties of the gas around the discontinuity are derived from spectra extracted 
from 0.5$^{\prime}$ wide partial annuli. We extract spectra from eight partial annuli 
confined by the pie shaped region shown in Fig.~\ref{fig:wdem-tempmap} with radii ranging from 
1.2--5.2$^{\prime}$ with respect to the central cD galaxy. In the two lower panels of Fig.~\ref{fig:profiles}, 
we show the obtained temperature and iron abundance profiles from this region. For comparison, we 
also show these values for similar regions on the opposite side of the core, which are shown in gray.
The position of the jump in count rate is indicated by the dashed vertical line.
The temperature profiles confirm the asymmetry between the South-Western and North-Eastern region
of the cluster. However, the profile does not show a jump at the radius of the surface-brightness 
discontinuity. Instead, they show a steady but asymmetric temperature gradient. Interestingly, 
the radius where the temperature profiles of the two opposite parts cross each other is 
$\sim$2.6$^{\prime}$, which is the same as the radius where the two surface-brightness profiles cross.
The most remarkable feature is the jump in iron abundance at 3.2$^{\prime}$. 
The abundance drops from $\sim$ 0.75 to $\sim$ 0.5 with respect to solar abundance units, which is 
consistent with the iron abundance values obtained from the maps. The area 
on the opposite side of the cluster, however, does not show a jump at that radius. 

\begin{figure}[t]
\includegraphics[width=\columnwidth]{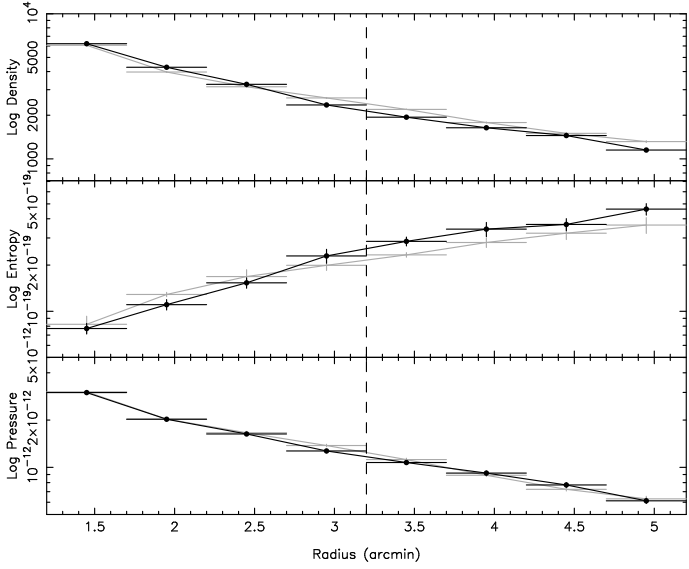}
\caption{Profiles of the derived density (m$^{-3}$, upper panel), entropy (N m$^{3}$, middle panel), and pressure 
(N m$^{-2}$, lower panel) from the cool South-Western region ({\it black}) and the opposite North-Eastern side 
({\it gray}) of Abell 2052. The dashed vertical line indicates the position
of the surface brightness discontinuity.}
\label{fig:profiles2}
\end{figure}

\subsection{Density, entropy, and pressure}

From the fits, we can calculate thermodynamic properties of the gas like entropy and pressure.
If the surface-brightness discontinuity we found in the South-West of the cluster is a shock, 
then we expect, for example, that the entropy increases after the shock has passed. For a cold front,
the contrary is expected. Then the entropy would be lower on the inner side of the discontinuity.
Since the entropy and pressure are calculated from the electron temperature and density of the 
gas, the density needs to be derived first. This is calculated using the normalisation 
($Y$) of the {\it wdem} component, which is given
by $Y = \int n_{\mathrm{e}} n_{\mathrm{H}} dV$, where $n_{\mathrm{H}}$ is the hydrogen density, $n_{\mathrm{e}}$ is equal to $1.2n_{\mathrm{H}}$ for Solar 
metalicity, and $V$ is the volume. In order to estimate the density for the annuli, 
we have to assume a size for the partial annulus along the line of sight to determine the volume. 
We choose this distance to be linearly proportional to the radius and in absolute value similar to 
the width of the partial annulus. The resulting density profile is shown in the top panel of 
Fig.~\ref{fig:profiles2}. The plot shows that the density is smoothly decreasing outwards. Around 
the discontinuity, the density is slightly lower than in a reference region on the other side of the
cluster, shown in gray. However, this plot does not show a discontinuity in density around
3$^{\prime}$.

\begin{figure}[t]
\includegraphics[width=\columnwidth]{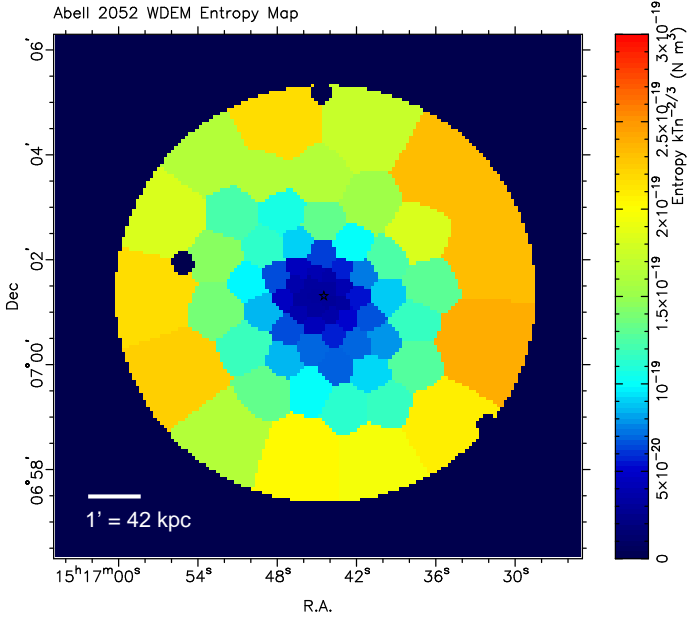}
\caption{Entropy map showing the (pseudo) entropy values, derived using the temperature and density
estimates obtained from the spectral fits. In the core, a star symbol marks the centre of the main cD galaxy.}
\label{fig:wdem-entropymap}
\end{figure}

Using this density profile, we can calculate the pseudo entropy following the equation 
$S = k T n_{\mathrm{e}}^{-2/3}$, where $S$ is the (pseudo) entropy, $k$ is the Boltzmann constant, 
$T$ is the temperature, and $n_{\mathrm{e}}$ is the electron density. The entropy profile is shown 
in the middle panel of Fig.~\ref{fig:profiles2}. A comparison between the profile around
the discontinuity and the profile on the other side of the cluster shows that the entropy
is also asymmetric. However, no strong entropy jump is seen around 3.2$^{\prime}$. The 
asymmetry is also present in the entropy map shown in Fig.~\ref{fig:wdem-entropymap}.
For these maps, the probed volume is estimated in a slightly different way compared to the 
profiles in Fig.~\ref{fig:profiles2}. We use the surface area of the bin in kpc$^2$ times a typical constant 
depth of 200 kpc. Then, of course, we neglect the slight dependence of the volume with radius. 
But since we are mainly interested in the relative differences, this dependence is not 
important for this discussion. The asymmetry shown in the entropy map appears to be much 
smoother than the structure seen in the temperature and iron abundance maps. The fact that
the entropy is lower in the inner region with respect to the discontinuity suggests
that the discontinuity is a cold front.

Finally, we can also derive the spatial profile of the pressure, which is shown in the
lower panel of Fig.~\ref{fig:profiles2}. The pressure is calculated using $P = n_{\mathrm{e}} k T$.
The profiles of both sides of the cluster are very similar
and smoothly decreasing outwards. Also the pressure map, which is not shown here, shows no
significant asymmetries. There are no discontinuities seen around 3.2$^{\prime}$, which suggest that 
the velocity of the front is low.   

\begin{table*}[!t]
\caption{Best-fit results for the multi-temperature fitting of the region with the high-alpha parameter.}
\begin{center}
\begin{tabular}{lccccccc}
\hline\hline
Parameter	& {\it wdem}		& {\it wdem}\tablefootmark{a}	& {\it gdem}-log	&{\it gdem}-lin & 4-temp\tablefootmark{b}	& 4-temp\tablefootmark{c}	& 1-temp \\
\hline
$kT_{\mathrm{max}}$ (keV)	
		& 3.69$\pm$0.05\tablefootmark{d}	& 3.36$\pm$0.06\tablefootmark{d}	& 2.06$\pm$0.02		& 2.16$\pm$0.03	& --		& 3.65$\pm$0.16	& --	\\
$kT_{\mathrm{mean}}$ (keV)
		& 2.25$\pm$0.04		& 2.22$\pm$0.04		& 	--		& --		& 2.28$\pm$0.03	& 2.25$\pm$0.12 & 1.482$\pm$0.011	\\
$\alpha$	& 4.0$\pm$0.8		& 1.14$\pm$0.05		&            -- 	&	     --	& --		& --		& --	\\
$\beta kT_{\mathrm{max}}$ (keV) 	
		& 0.61$\pm$0.02		& --			& --			& --		& --		& --		& --	\\
$\sigma_T$ (keV)& 	     -- 	& --			& 0.338$\pm$0.007\tablefootmark{e}	& 1.34$\pm$0.05	& --		& --		& --	\\
Fe		& 1.20$\pm$0.03 	& 1.27$\pm$0.03		& 1.14$\pm$0.02		& 1.22$\pm$0.04	& 1.27$\pm$0.04	& 1.21$\pm$0.05	& 0.438$\pm$0.015	\\
C-stat/d.o.f.   & 1207 / 940		& 1270 / 941		& 1199 / 941		& 1262 / 941	& 1163 / 940	& 1160 / 939	&  2795 / 942	\\
\hline
\end{tabular}\\
\end{center}
\tablefoot{
\tablefoottext{a}{In this {\it wdem} model the low-temperature cut-off ($\beta$) is fixed to 0.1.}
\tablefoottext{b}{4-temperature model with the temperatures fixed to 0.5, 1, 2, and 4 keV.}
\tablefoottext{c}{4-temperature model with the maximum temperature free and other temperatures coupled with factors 0.5, 0.25, and 0.125.}
\tablefoottext{d}{This is the maximum temperature of the DEM distribution kT$_{\mathrm{max}}$.}
\tablefoottext{e}{This is the $\sigma_T$ of the Gaussian DEM distribution on a logarithmic grid.}
}
\label{tab:multit}
\end{table*}

\begin{figure*}[!t]
\sidecaption
\includegraphics[width=12cm]{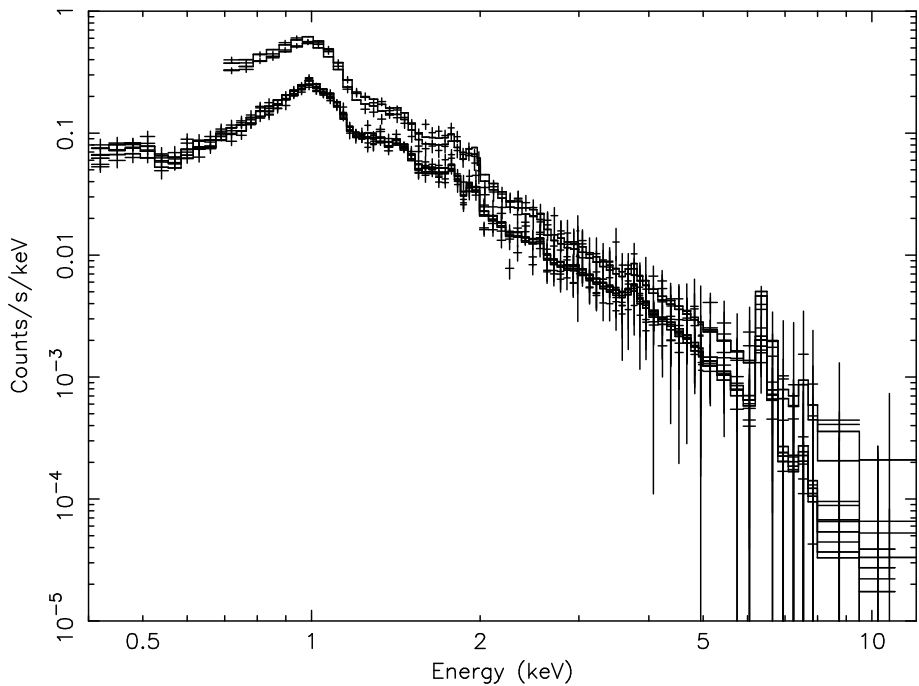}
\caption{EPIC MOS and pn spectra for AO1 and AO4 extracted from the North-Western region, which shows the
low-temperature cut-off of $\sim$0.5 keV. The data and best-fit model both represent the total spectrum including 
all the cosmic and instrumental background components. The models start diverging above 8 keV, because the 
instrumental background starts to dominate in this energy band.}
\label{fig:spectrum}
\end{figure*}

\subsection{Multi-temperature regions}

Since we used a multi-temperature {\it wdem} model to fit the spectra from the maps, there is also an estimate for 
the slope of the DEM distribution $\alpha$. In Fig.~\ref{fig:wdem-alphamap}, we show a map of $\alpha$ 
with $\beta$ fixed to 0.1. 
A value near zero means that the DEM distribution is very peaked (near single-temperature), and a value
significantly larger than zero means that there is a significant contribution of low-temperature gas.
Surprisingly, there is a small region in the cluster with a large value for $\alpha$, which is located
just to the North-Western side of the central galaxy. This region is indicated with an arrow in 
Fig.~\ref{fig:wdem-alphamap}. The maximum value for $\alpha$ is 1.06$\pm$0.09, while the 
values in the immediate surroundings are typically 0.3. The large value suggests a large contribution of a
cool component. 

These bins that have an unusually high $\alpha$ value
are associated with a very interesting multi-temperature region. Therefore, we extract a spectrum from
a circular region centered on the bin with the high $\alpha$ value. The resulting spectra for 
EPIC MOS and pn are shown in Fig.~\ref{fig:spectrum}. The radius of the extraction 
region is 15$^{\prime\prime}$. We attempt to put constraints on the DEM distribution by fitting
six empirical DEM distributions to the spectrum, which we explain in Section~\ref{sec:models}.
The results of the fits are listed in Table~\ref{tab:multit}. Although there are some variations, 
the best-fit C-statistic values for the multi-temperature fits are very similar. The iron abundances that we 
derive from the multi-temperature models are typically around 1.25 solar, except for the logarithmic 
Gaussian DEM model, which gives a lower iron abundance of 1.14$\pm$0.02. In general, the derived mean 
temperatures and iron abundances are almost the same regardless of the used multi-temperature model. However,
this does not hold for a single-temperature model. A single-temperature fit to this region results in 
an unacceptable C-stat value of 2795 / 942 d.o.f. The temperature and iron abundance values are 
unrealistically low compared to the multi-temperature results. Therefore, we ignore this model 
in the discussion about the temperature structure of this region.

\begin{figure}[t]
\includegraphics[width=\columnwidth]{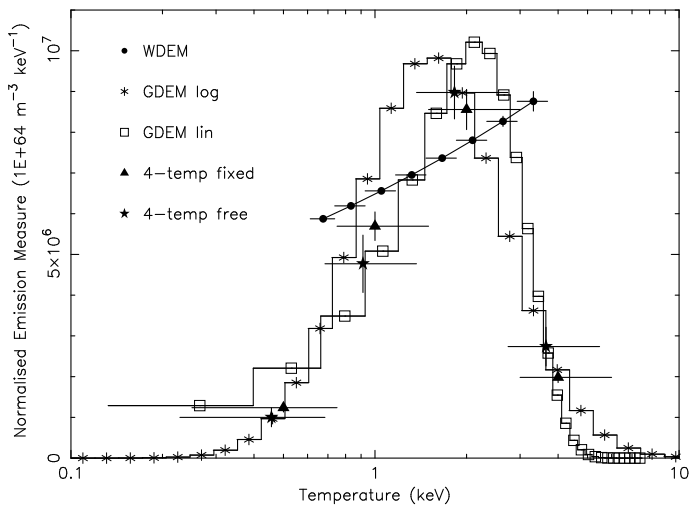}
\caption{Overview of the best-fit DEM distributions fitted to the North-Western multi-temperature 
region. The emission measure is normalised based on the width of the temperature bin. For the
4-temperature models, the bin width is somewhat arbitrary. We choose asymmetric bin sizes to form 
a continuous region from 0.5 up to 4 keV.}
\label{fig:dem}
\end{figure}

The multi-temperature distributions that we fit are best compared by plotting them. In Fig.~\ref{fig:dem},
we plot the DEM distributions and normalisations of five multi-temperature models. In this plot, the 
emission measures are normalised using the bin width to be able to compare them directly. 
Since the best-fit models have similar C-statistic values, the exact shape of the DEM distribution
is uncertain. However, they all show a similar trend. The peak temperature of the distribution 
is found around temperatures of 2 and 3 keV. Above 3 keV, the contribution of high temperatures
drops rapidly. Below 2 keV all models show a significant contribution of cool gas even down to 
0.5 keV. Interestingly, the four-temperature models show an emission measure distribution that is comparable to the 
Gaussian DEM models.

Using the 4-temperature model with the fixed temperatures, we estimate the volume-filling fractions of the temperature
components if we assume that the gas is in pressure equilibrium. For component $i$ of $n$ 
temperature components, the volume filling fraction ($f_i$) for a multi-phase gas is given by \citep{sanders2004}:
\begin{equation}
f_i = \frac{Y_i T_{i}^{2}}{\displaystyle \sum_{j=0}^n Y_j T_{j}^{2}},
\label{eq:volfil}
\end{equation}
where $Y_i$ is the emission measure and $T_i$ the temperature of the $i$-th component. 
For the 0.5 keV component of the 4-temperature model with the fixed $kT_{\mathrm{max}}$ of 4.0 keV, we 
find a volume filling fraction $f$ of (1.02$\pm$0.09) $\times$ 10$^{-3}$ and for the 1.0 keV component 
we find (2.83$\pm$0.17) $\times$ 10$^{-2}$. In other clusters, these fractions are typically of the 
order of 10$^{-3}$ and 10$^{-2}$ for temperatures of 0.5 and 1.0 keV, respectively. Fractions of this order
have been measured in, for example, \object{Abell 3281} \citep{sanders2010} and \object{M87} \citep{werner2010}. 

We checked the spectrum for the presence of a cool 0.2 keV component and find a one sigma upper limit 
for the emission measure of $Y <$ 2.5$\times$10$^{69}$ m$^{-3}$. In principle, this measurement
may be slightly biased due to the Galactic background components that have similar temperatures.
However, these background components have an order of magnitude lower emission 
measure ($Y \sim$ 10$^{68}$ m$^{-3}$) if they were placed at the cluster redshift. Since we fix the
background to the emission measures we found in the outer parts of the cluster, the upper limit 
is robust.

\section{Discussion}

In the maps derived from XMM-Newton EPIC data of \object{Abell 2052}, we find two striking features in 
addition to what was already known about the cluster structure from detailed Chandra observations 
\citep{blanton2001,blanton2003,blanton2009}. To the South-West of the cluster there is a large region with
a relatively low entropy and temperature, but with a higher iron abundance. The Southern border
of the region is formed by a surface brightness discontinuity. Just North-West of the central galaxy, 
we find a smaller region, which shows an unusually high contribution of cool (0.5 keV) gas. We 
focus the discussion on these two findings.  

\subsection{The origin of the Southern asymmetry}

Asymmetries in the temperature, entropy and abundance maps like we see to the South-West of
the core of \object{Abell 2052} are fairly common in clusters of galaxies. Usually, these low-temperature 
regions, or cold fronts, are confined by a sharp surface brightness discontinuity \citep[see][for a review]{markevitch2007}.
For Abell 2052, \citet{blanton2009} already describe the elliptical shape of inner the core in the 
North-South direction. But just outside the region with the X-ray cavities, the core shows 
ellipticity in the NE-SW direction with an enhanced surface brightness area in the South-West.
This NE-SW ellipticity is seen in both XMM-Newton and Chandra images. A sharp discontinuity,
however, is not seen directly in the current images. By extracting a radial profile for a pie 
shaped region in the South, we do see a discontinuity in the mean count rate at an angular distance
of 3.2$^{\prime}$ from the central cD galaxy. 

The ellipticity of the low-temperature region, the relatively weak discontinuity at
the edge, and no evidence for structure in the pressure map or pressure profile suggest that the asymmetry 
is not due to a recent large merger event. Also the cooling core is not disrupted, which means
that the cause of the disruption was either weak or a long time ago. The fact that we see 
only a significant jump in surface brightness and Fe abundance across the discontinuity 
most likely indicates gas sloshing \citep{markevitch2001}. In this case, a small merger event or 
a larger merger in the distant past caused the dark matter and the associated cool gas to 
oscillate in the gravitational potential of the cluster. We only observe a gradual 
temperature, density, entropy, or pressure gradient across the cold front, and the entropy on the inside of the 
discontinuity is lower than on the outside. This means that the velocity of the front must be subsonic,
and the discontinuity is not a shock. 
 
Contrary to the lack of significant thermodynamic discontinuities found in the profiles, the 
iron abundance jumps down roughly 0.2 solar units across the front. Numerical simulations show 
that sloshing motions can indeed cause discontinuities in abundance profiles \citep{ascasibar2006}. 
This jump in iron abundance suggests that the sloshing motions are a mechanism to transport  
low-entropy metal-rich gas to the outer parts of the cluster. Recently, the
same effect has also been observed in \object{M87} \citep{simionescu2010}. The fact that metal transport
is observed in these deep observations of clusters of galaxies suggests that gas sloshing is 
generally a very important and a dominating effect in the transport of metals to 
the outer parts of clusters. 

It is interesting that the discontinuity in the iron abundance is very sharp, while the temperature 
profile changes more gradually. This has not only been observed in Abell 2052, but, for example, also
in \object{M87} \citep{simionescu2010}. The general shape of the surface brightness, temperature, and iron abundance
profiles in Abell 2052 look remarkably similar to those observed in M87. The shape suggests that 
conduction may play a larger role than previously thought. Earlier observations have shown that 
conduction should be suppressed across cold fronts \citep{ettori2000,markevitch2007} likely due 
to the presence of magnetic fields parallel to the front. However, when a cold front ages, thermal 
conduction between the cool and hot gas may gradually smooth the temperature jump without affecting 
the jump in metalicity. Therefore, conduction appears to be more important than, for example, mixing, 
which would also smooth the metalicity jump. 

\begin{figure}
\includegraphics[width=\columnwidth]{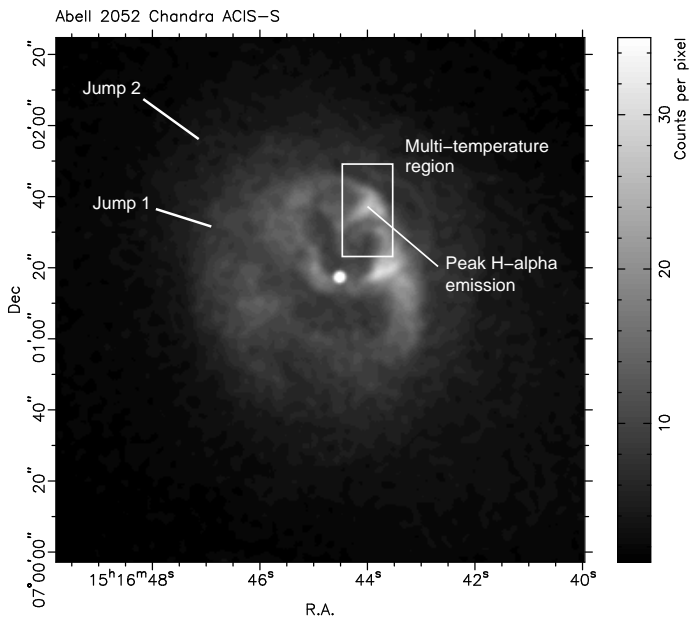}
\caption{Chandra ACIS-S image of Abell 2052. The bright point source in the centre is 
associated with the 3C 317 radio source. The image was smoothed using a Gaussian with a $\sigma$ of 
2 pixels. The box in the image indicates the region with a high {\it wdem} $\alpha$ parameter. In addition,
we indicate a local peak of H$\alpha$+\ion{N}{II} emission \citep{baum1988}. On the North-Eastern side, we
indicate to surface brightness jumps identified by \citet{blanton2009}.}
\label{fig:acis}
\end{figure}

In numerical simulations of cold fronts, often more than one cold front is seen on both
sides of the cluster \citep[e.g.][]{tittley2005}. If there are more then one, they are expected
to alternate on an axis through the centre of the cluster. In Abell 2052, the counterpart of the front
at 3.2$^{\prime}$ may lie North-East of the centre. \citet{blanton2009}
identified two jumps in surface brightness in that area using a deep Chandra image. The locations
of these jumps are indicated in Fig.~\ref{fig:acis}. They are located at 45$^{\prime\prime}$ 
and 67$^{\prime\prime}$ from the cluster centre. The 67$^{\prime\prime}$ jump may be the counterpart of
the cold front at 3.2$^{\prime}$. If we look back to Fig.~\ref{fig:profiles},
then the iron abundance profile of the NE side of the central galaxy shows a jump around 1.7$^{\prime}$.
Probably, this jump is a bit too far from the jumps identified in the Chandra data to be associated
with the counterpart of the cold front we find, although the spatial resolution of our iron abundance 
profile is relatively low. In addition, this North-Eastern region may be affected by the AGN 
activity in the core of the cluster. However, the position of the NE jumps in surface brightness
and iron abundance appear to support the core oscillation interpretation. 

\subsection{Cool multi-temperature gas and H$\alpha$ emission}

In the map showing the slope of the DEM distribution, we find a small region North-West 
of the central galaxy that shows a relatively 
large contribution of cool gas. This region, with approximate dimensions of 10$\times$20 kpc,
is indicated on an archival Chandra ACIS-S image of Abell 2052 in Fig.~\ref{fig:acis}. It 
coincides with a relatively bright spur of X-ray emission \citep{blanton2001}.
We fit the multi-temperature region North-West of the central galaxy with several multi-temperature
models featuring different assumptions for the DEM distribution. From the C-statistic values it is 
clear that single-temperature models are inadequate to describe the observed spectrum, especially
the temperature sensitive Fe-L complex. The underlying temperature structure is however difficult
to constrain with the current instrumentation. \citet{kaastra2004b} already show that spectra 
calculated from different DEM distributions with the same emission weighted average temperature 
and emission measure are almost indistinguishable at low spectral resolution. It is therefore 
not very surprising that we find similar C-statistic values for the {\it wdem}, {\it gdem}, and 
4-temperature fits.

Although the exact shape of the DEM distribution cannot be derived, the global trends are quite
robust. We can conclude from Fig.~\ref{fig:dem} that there is a significant amount of cool gas
with temperatures of about 0.5--1.0 keV. The 1 keV component has about half of the emission measure
with respect to the peak component of about 2.0 keV and the 0.5 keV component still has an emission
measure between 10--20\% of the peak component. Like in other clusters \citep[e.g.][]{sanders2010},
there appears to be a large drop in emission measure below a temperature of 0.5 keV. With the 
{\it wdem} model, we fit this low-temperature cut-off $\beta kT$ and find values consistent with  
temperatures in the range of 0.5--0.6 keV. In addition, we find a one $\sigma$ upper limit for the emission measure
for a 0.2 keV component of $Y <$ 2.5$\times$10$^{69}$ m$^{-3}$, which is $<$ 1.1\% of the total emissivity.
For a cooling-flow model with a mass deposition rate of 8 M$_{\odot}$/yr, based on 
the 0.5 and 1.0 keV emission, an emission measure of $Y \sim$ 10$^{70}$ m$^{-3}$
would be expected. This suggests that a physical mechanism should be at work at temperatures 
lower than 0.5 keV to prevent the gas from cooling radiatively. 

Comparing this low-temperature cut-off directly with other clusters is not trivial, because for 
spectra with low-statistics, $\alpha$ and $\beta$ are correlated. In previous
papers the $\beta$ parameter in the {\it wdem} model was often fixed to a low value (e.g. 0.1$kT_{\mathrm{max}}$).
A large contribution of cool gas is then indicated by a large value of $\alpha$, which means a rather 
flat temperature distribution down to low temperatures.  
If we compare the value for $\alpha$ derived from our {\it wdem} fit to values found in the core
regions of other clusters, then our value of 1.14$\pm$0.05 is relatively high. In the cluster
sample of \citet{kaastra2004}, the values in the inner annuli are typically varying between 0.2--0.8,
with the exception of Abell 1835 which shows an $\alpha$ value of 1.3$\pm$0.5. Another example of
a high $\alpha$ value of 1.1$\pm$0.3 was found by \citet{deplaa2004} in the central region of Abell 478,
which also shows evidence of central AGN activity, like Abell 2052. The error bars on these measurements, 
however, are relatively large and the extraction regions of the spectra were not selected based on an 
$\alpha$ map. More recently, high $\alpha$ regions were found just to the east of M87 \citep{simionescu2008}.
There are just a few bins in the $\alpha$ map that are larger than 1, which is similar to the situation in 
Abell 2052. Since \citet{simionescu2008} leave the low-temperature cut-off ($\beta$) free, directly 
comparing absolute values of $\alpha$ is not possible. Relative differences in maps should, however, not 
be affected. 

Detailed multi-wavelength studies of \object{M87} show that these coolest regions are associated with H$\alpha$ 
filaments \citep{young2002,sparks2004,werner2010}. Similar 
regions were also found in other bright cool-core clusters or galaxies like, for example,  
\object{Perseus} \citep{fabian2003b,sanders2007}, \object{Centaurus} \citep{crawford2005}, and \object{2A 0335+096} \citep{sanders2009}. 
In the multi-temperature region of Abell 2052, there is also a correlation between high $\alpha$ values
and H$\alpha$ emission. H$\alpha$ data \citep{baum1988} overlaid on Chandra images \citep{blanton2001} 
show that the region where we find the high $\alpha$ values are also showing enhanced H$\alpha$+\ion{N}{II} 
emission. The peak of the H$\alpha$+\ion{N}{II} emission is exactly centered on the region with the 
enhanced cool components, within the resolution of our map. The H$\alpha$ data of \citet{baum1988} 
suggest that the blob of cool ($T \sim 10^{4}$ K) gas has been uplifted or pushed up by the Northern bubble, because
there is a small filament of H$\alpha$ emission extending from the central galaxy toward the North-Western  
blob. The Chandra image suggests that the region may be squeezed by the Northern large bubble, therefore increasing 
its density, and shorten its cooling time. Assuming pressure equilibrium, the radiative cooling time of 
the 0.5 keV gas in this region would be $\sim$ 4$\times$10$^{6}$ yr. At this position, the blob is able
to cool efficiently without being subsequently accreted or heated by the AGN. The only feedback mechanism
that could heat the gas would be supernovae originating from star forming induced by continued gas cooling.
Star formation is detected in optical and UV-images of the central cD galaxy, but it has not been 
reported at the position of this H$\alpha$ spur \citep{martel2002,blanton2003}. In order to constrain 
the star formation rate due to gas cooling at the position of the spur, we searched for UV emission of  
young stars using data from the Optical Monitor (OM) aboard XMM-Newton. In a 29 ks OM exposure using the 
UVW1 filter, we also do not detect enhanced UV emission at this position. From the data, we derive a one 
sigma upper limit for the UV luminosity of a possible star forming region of $L$ = 1.21$\times$10$^{18}$ W Hz$^{-1}$.
This would roughly correspond to a star formation rate of 1.7$\times$10$^{-3}$ M$_{\odot}$/yr \citep{kennicutt1998},
which means that the gas does not continue cooling to star-forming temperatures. 

It is still unclear how the brightness of the optical H$\alpha$ emission can be explained. For the small sample of 
\citet{sanders2010}, the H$\alpha$ luminosities in these regions are comparable or even higher than the 
missing X-ray luminosity below 0.5 keV ($\sim$ 10$^{35}$ W s$^{-1}$). Most likely, the
H$\alpha$ emission is concentrated in thin filaments \citep{sharma2010} embedded in the hotter 0.5 keV gas.
The volume filling fraction of 1.1 $\times$ 10$^{-3}$ for the 0.5 keV component would support the clumpy
or filamentary nature of this gas. Since gas cooler than 0.5 keV is not detected in X-rays, the 0.5 keV
gas is probably cooling non-radiatively. \citet{fabian2002} suggest that heat is transferred from the 
0.5 keV gas to the H$\alpha$ emitting gas through mixing or conduction. Since the H$\alpha$ emitting gas 
is probably magnetised, conduction and mixing may not be fully responsible for heating the H$\alpha$ gas. 
\citet{fabian2008} and \citet{ferland2009} show that magnetic fields in these filaments can be relatively 
strong. 

Small MHD shocks or waves may also play a role in mixing the coolest X-ray and cold H$\alpha$ gas 
phases. \citet{werner2010} propose that passing shock waves can accelerate the ICM past the much denser
H$\alpha$ filaments, which induces shearing and subsequently Kelvin-Helmholtz instabilities that mix the
two phases. The hot ICM electrons then ionize and heat the cool gas, which induces H$\alpha$ emission. 
Shocks and sound waves have been seen in recent Chandra observations of Abell 2052 \citep{blanton2009}. 
Chandra images show a lot of smaller shock fronts or sound waves around the edges of the bubbles, and also 
just in front of the blob of H$\alpha$ emission. \citet{werner2010} base this hypothesis on the fact that most
H$\alpha$ regions are seen in the downstream regions of a shock. Therefore, this mechanism could explain 
why the X-ray gas would cool non-radiatively below 0.5 keV. Mixing and shock heating may also explain why 
the gas does not cool all the way to star-forming temperatures. Since observations of \object{M87}, for example, 
have shown that the H$\alpha$ filaments contain dust \citep{sparks1993}, the shocks should just gently 
heat the gas and not evaporate the dust. The multi-temperature region of \object{Abell 2052} appears to be fully 
consistent with the H$\alpha$ and 0.5 keV gas interactions as seen in \object{M87}. However, in a recent  
H$\alpha$ survey of 23 clusters \citep{mcdonald2010}, \object{Abell 2052} is one of few objects where the 
NUV/H$\alpha$ flux ratio is consistent with shock heating. Therefore, this region may prove to be a very 
interesting special case. More detailed H$\alpha$, CO, UV measurements are necessary to unravel the heating 
and feedback mechanisms operating in this filament.

\section{Conclusions}

Using a deep XMM-Newton observation (95 ks), we have derived 2D maps of the core of the cluster Abell 2052.
In the maps, we discover a cold front at a distance of $\sim$ 130 kpc in the South-Western direction from 
the central cD galaxy. Close to the cD galaxy in the North-Western direction, we find a multi-temperature 
region with cool X-ray emitting gas as low as 0.5 keV. From a careful analysis of these regions, 
we conclude that:  

\begin{itemize}
\item We find a small local cooling-flow region NW of the cD galaxy. Most likely, this gas has been uplifted
or pushed away from the core and squeezed by a nearby bubble, where it can cool efficiently and relatively 
undisturbed.  
\item In the cooling-flow region NW of the cD galaxy, we do not detect gas below 0.5 keV. Although we 
cannot constrain the shape of the Differential Emission Measure distribution, the upper limit for gas 
around 0.2 keV is robust and lower than the emission measure expected from cooling-flow models.
\item The lack of lines from gas below 0.5 keV may be explained by the presence of H$\alpha$ filaments. 
Shock induced mixing between the two phases may cause the 0.5 keV gas to cool non-radiatively.
\item We find significant jumps in surface brightness and iron abundance across the cold front in the 
South-Western part of the cluster, but no indications for a high Mach number. Therefore, it is a cold 
front and most likely the result of gas sloshing.
\item The sharp jump in iron abundance across the cold front suggests that sloshing is at least partly 
responsible for transporting metals from the core region to the outer parts of the cluster. In addition,
the smooth temperature profile in the same area as the sharp iron jump suggests that conduction is 
responsible for smoothing the temperature gradient instead of mixing.
\end{itemize}

\begin{acknowledgements}
This work is based on observations obtained with XMM-Newton, an ESA science mission
with instruments and contributions directly funded by ESA Member States and
NASA. The Netherlands Institute for Space Research (SRON) is supported financially
by NWO, the Netherlands Organisation for Scientific Research. 
N. Werner and A. Simionescu were supported by the National Aeronautics and Space Administration through 
Chandra/Einstein Postdoctoral Fellowship Award Numbers PF8-90056 and PF9-00070 issued 
by the Chandra X-ray Observatory Center, which is operated by the Smithsonian Astrophysical 
Observatory for and on behalf of the National Aeronautics and Space Administration under contract NAS8-03060.
\end{acknowledgements}

\bibliographystyle{aa}
\bibliography{clusters}

\end{document}